\documentclass[aps,showpacs,preprint,graphicx]{revtex4}
\usepackage[dvips]{graphicx}
\begin{document}
\title{Plasma dispersion in fractional-dimensional space} 
\author{K. M. Mohapatra}
\affiliation{Department of Physics, North Odisa University, Baripada, Odisha, India.}
\author{Dr. B. K. Panda}
\affiliation{Department of Physics, Ravenshaw University, Cuttack, Orissa, India.}
\begin{abstract}
The dielectric function for electron gas with parabolic energy bands 
is derived in a fractional dimensional space. 
The static response function shows a good dimensional dependence. 
The plasma frequencies are obtained 
from the roots of the dielectric functions. The plasma dispersion 
shows strong dimensional dependence. It is found that the 
plasma frequencies in the low dimensional 
systems are  strongly dependent on the wave vector. It is weakly dependent 
in the three dimensional system and has a finite value at zero wave vector.
\end{abstract}
\maketitle
\section{Introduction}
When the well width of a quantum well (QW) is extremely 
narrow and its barrier height 
that causes the in-plane confinement is 
infinite, the QW shows two-dimensional (2D)
electronic and optical properties. The infinitely 
wide QW exhibits the three-dimensional (3D)
bulk properties of the well material\cite{Matos}.   
The electronic and optical properties in a QW with finite barrier height 
and narrow well width show 3D behavior of the barrier material.  
It happens since the envelope functions for electrons
and holes spread into the barrier region partially 
restoring the 3D characteristics of the system. 
On the other hand, the electronic and optical properties 
in a finite QW with sufficiently wide well width show 3D characteristics 
of the well material.   
Consequently the QW with finite well 
width and barrier height shows 
the fractional dimensional behavior which is 
somewhere in between 2D and 3D. This
has been demonstrated by Ishida\cite{Ishida} in the 
calculation of plasma dispersion in 
a superlattice. The same behavior has also been demonstrated 
in the calculation of exciton\cite{Jai} and polaron\cite{Smondyrev} 
ground state properties.

The anisotropic interactions in an anisotropic solid
are treated as ones in an isotropic
fractional dimensional space, where the dimension is
determined by the degree of anisotropy\cite{He}.
Thus only a single parameter known as the degree of
dimensionality $(\alpha)$ is
needed to describe the system.
In the quantum well structures the width of the QW
can also serve for determining $\alpha$
of the system.
The fractional dimensional $\alpha D$ space is not Euclidean 
space, it is spectroscopic dimension 
which is observed\cite{Bak}. 
The $(\alpha$D) space is not a vector space and
the coordinates in this space are termed as
{\em pseudo coordinates}\cite{Stillinger}.

The advantage of the $\alpha$D space approach over the conventional method
for calculating different electronic and optical properties in the 
low dimensional systems is that 
it is easier to apply this method. For example, the $\alpha D$ space approach
has been successfully employed
to calculate exciton binding energy in QWs in an analytic
method\cite{He1,Matos1}, while the
conventional method needs involved numerical calculations\cite{Jai}.
Similarly the polaron properties in the $\alpha D$ space have been studied
in a simple method\cite{Polaron} whereas the conventional method needs quite a bit 
of computational effort\cite{Smondyrev}.  
The technique has also been used to study biexcitons\cite{Biex1,Biex2,Biex3},
magnetoexciton\cite{Magnet1,Magnet2}, exciton-exciton interaction\cite{Exex},
exciton-phonon interaction\cite{Exph},
Stark shift of exciton complexes in weak electric field\cite{Stark},
refractive index\cite{Tanguy}, impurity and
donor states\cite{Imp1,Harrison,Imp2},
Pauli blocking effect\cite{Pauli}, 
exciton-phonon interaction\cite{Exphn},
exciton-polaron interaction\cite{Expol} and magnetopolaron\cite{Magpol}.
The absorption spectra in a quantum wire shows the fractional 
dimensional space behavior with the dimension of the
system lies between 1 and 3  depending on the size of the system\cite{Karlsson}. 

Several properties of the charged boson system have been studied
in the $\alpha D$ space using the Singwi-Tosi-Land-Sj\"olander (STLS)
method\cite{Boson}.
The Luttinger liquid\cite{Castelliani} and the breakdown of Fermi liquid due to long range 
interaction\cite{Wirefd} in the 
fractional dimensional space with the dimension between 1 and 2 have been studied.
In the fractional dimensional space the  plasma frequencies in the
long-wavelength limit have been derived from the real part of the
dielectric function both in the quantum and classical
limits\cite{Panda}. However,
the full treatment of the
dielectric function in the $\alpha D$ space for finding plasma
frequency has not been carried out for Fermi gas. The present paper aims to fill up 
this gap and study the correlation energy.  

\section{Dielectric function} 

In the $\alpha D$ space, the dielectric function with the wave vector
$q$ and frequency $\omega$
of the external charge is defined as\cite{Vignale}
\begin{equation}
\epsilon_{\alpha D}(q,\omega)=1-v_{\alpha D}(q)\chi^{0}(q,\omega),
\end{equation}
where $v_{\alpha D}(q)$ is the Fourier transform of Coulomb potential 
$e^{2}/\epsilon_{\infty}r$
in $\alpha$D space and 
$\chi^{0}_{\alpha D}(q,\omega)$ is the irreducible polarizability function. 
The expression for $v_{\alpha D}(q)$
is given as\cite{Stillinger},
\begin{equation}
v_{\alpha D}(q)=\frac{(4\pi)^{\frac{\alpha-1}{2}}
\Gamma\biggl(\frac{\alpha-1}{2}\biggr)e^{2}}
{q^{\alpha-1}}\label{eq:cq},
\end{equation}
where $\Gamma(x)$ is the Euler gamma function. 

The irreducible polarizability function is defined as\cite{Vignale}
\begin{equation}
\chi^{0}_{\alpha D}(q,\omega)=\frac{2}{V_{\alpha D}}
\sum_{\bf k}\frac{f({\bf k})-
f({\bf k}+{\bf q})}
{E_{\bf k}-E_{{\bf k}+{\bf q}}+\hbar\omega+i\epsilon}\label{eq:pol1},
\end{equation}
where $f({\bf k})$ is the Fermi-Dirac distribution function,  
$V_{\alpha D}$ is the volume in $\alpha$D space and $\epsilon\rightarrow 0^{+}$. 
Rearranging Eq.(\ref{eq:pol1}), we find
\begin{equation}
\chi^{0}_{\alpha D}(q,\omega)=-\frac{2}{V_{\alpha D}}\sum_{\bf k}f({\bf k})
\left[\frac{1}{E_{{\bf k}+{\bf q}}-
E_{\bf k}-\hbar\omega-i\epsilon}
+\frac{1}{E_{\bf k-{\bf q}}-E_{{\bf k}}+
\hbar\omega+i\epsilon}\right]\label{eq:pol}
\end{equation}
We consider the zero-temperature limit and the parabolic energy dispersion,  
$E_{\bf k}=\hbar^2k^2/2m^{\ast}$ where $m^{\ast}$ is the effective mass
of electron. The summation over {\bf k} in the $\alpha D$ space approach is
transferred into integration over $k$ and $\theta$ as
\begin{equation}
\sum_{\bf k}=\frac{V_{\alpha D}}{(2\pi)^{\alpha}}
\frac{2\pi^{\frac{\alpha-1}{2}}}{\Gamma\biggl(\frac{\alpha-1}{2}\biggr)}
\int^{k_F}_{0}k^{\alpha-1}dk\int^{\pi}_{0}
\sin^{\alpha-2}\theta d\theta\label{eq:int}
\end{equation}

In the $\alpha D$ space, the Fermi momentum $k_{F}$ is related to $r_{s}$ as
\begin{equation}
k_{F}r_{s}a_{B}=\beta_{\alpha}\label{eq:rs},
\end{equation}
where $a_{B}$ is the Bohr's radius and $\beta_{\alpha}=[2^{d-1}\{\Gamma(1+d/2)\}^2]^{1/\alpha}$.
\begin{eqnarray}
\chi^{0}_{\alpha D}(q,\omega)&=&-
\frac{2^{2-\alpha}}
{\pi^{\frac{\alpha+1}{2}}\Gamma\biggl(\frac{\alpha-1}{2}\biggr)}
\int^{k_F}_{0}k^{\alpha-1}dk \int^{\pi}_{0}
\sin^{\alpha-2}\theta d\theta\nonumber\\
& & \left[\frac{1}{E_{q}+\hbar^{2}kq\cos\theta/m^{\ast}-
\hbar\omega-i\epsilon}+
\frac{1}{E_{q}-\hbar^{2}kq\cos\theta/m^{\ast}+
\hbar\omega+i\epsilon}\right]\label{eq:pol2}
\end{eqnarray}

We have the identity
\begin{equation}
\frac{1}{x\pm i\epsilon}=P\biggl[\frac{1}{x}\biggr]
\mp i\pi\delta(x)\label{eq:identity},
\end{equation}
where $P[1/x]$ is principal part of $1/x$ and $\delta(x)$ is the Dirac delta function.

\subsection{Real part of the dielectric function}

Using Eq.(\ref{eq:identity}) in 
Eq.(\ref{eq:pol2}), we find
\begin{eqnarray}
Re[\chi^{0}_{\alpha D}(q,\omega)]&&=-
\frac{2^{2-\alpha}}
{\pi^{\frac{\alpha+1}{2}}\Gamma\biggl(\frac{\alpha-1}{2}\biggr)}
\int^{k_F}_{0}k^{\alpha-1}dk \int^{\pi}_{0}
\sin^{\alpha-2}\theta d\theta\nonumber\\
&& P\Biggl[\frac{1}{E_{q}+\hbar^{2}kq\cos\theta/m^{\ast}-
\hbar\omega}+
\frac{1}{E_{q}-\hbar^{2}kq\cos\theta/m^{\ast}+
\hbar\omega}\Biggr]\label{eq:pol}
\end{eqnarray}

The first and second integrals in Eq.(\ref{eq:pol}) diverge when 
$\theta=\theta_{-}$ and $\theta=\theta_{+}$, respectively where
$\theta_{\pm}=\arccos[m^{\ast}(\hbar\omega\pm E_{q})/\hbar^2kq$.

\begin{eqnarray}
Re[\chi^{0}_{\alpha D}(q,\omega)]&&=-
\frac{2^{2-\alpha}}
{\pi^{\frac{\alpha+1}{2}}\Gamma\biggl(\frac{\alpha-1}{2}\biggr)}
\int^{k_F}_{0}k^{\alpha-1}dk\nonumber\\
&& \left[\int^{\pi}_{0}d\theta
\frac{\left(\sin^{\alpha-2}\theta-\sin^{\alpha-2}\theta_-\right)}
{E_q-\hbar\omega+\hbar^2kq\cos\theta/m^{\ast}} + \int^{\pi}_{0}d\theta
\frac{\left(\sin^{\alpha-2}\theta-\sin^{\alpha-2}\theta_+\right)}
{E_q+\hbar\omega-\hbar^2kq\cos\theta)/m^{\ast}}\right]
\end{eqnarray}
Carrying out the $\theta$ integration analytically, we find 
\begin{eqnarray}
Re[\chi^{0}_{\alpha D}(q,\omega)]&=&-
\frac{2^{2-\alpha}}{\pi^{\frac{\alpha}{2}}
\Gamma\biggl(\frac{\alpha}{2}\biggr)}
\int^{k_{F}}_{0}k^{\alpha-1}dk\nonumber\\
& & \Biggl[\frac{{_2F_1}\biggl(1,\frac{\alpha-1}{2},\alpha-1;
\frac{2\hbar^2kq/m^{\ast}}{(E_q-\hbar\omega)
-\hbar^2kq/m^{\ast}}\biggr)}{(E_q-\hbar\omega)
-\hbar^2kq/m^{\ast}}
+\frac{\pi\sin^{\alpha-2}\theta_-}
{(E_q-\hbar\omega)^2-(\hbar^2kq/m^{\ast})^2}\nonumber \\
&& +\frac{{_2F_1}\biggl(1,\frac{\alpha-1}{2},\alpha-1;
\frac{2\hbar^2kq/m^{\ast}}{(E_q+\hbar\omega)
+\hbar^2kq/m^{\ast}}\biggr)}{(E_q+\hbar\omega)
+\hbar^2kq/m^{\ast}}
+\frac{\pi\sin^{\alpha-2}\theta_+}
{(E_q+\hbar\omega)^2-(\hbar^2kq/m^{\ast})^2}\Biggr],
\end{eqnarray}
where ${_2F_1}$ is the Gauss Hypergeometric function. 
The real part of the susceptibility 
$\chi^{0}(q,\omega)$ does not contain any divergent part in the range
$\hbar v_{F}q-E_{q}<\hbar\omega<\hbar v_{F}q+E_{q}$. In this range
the real part of the susceptibility is obtained as
\begin{eqnarray}
Re[\chi^{0}_{\alpha D}(q,\omega)]&&=-
\frac{2^{2-\alpha}k^{\alpha}_{F}}{\pi^{\frac{\alpha}{2}}
\Gamma\biggl(\frac{\alpha-1}{2}\biggr)}
\sum^{\infty}_{m=0}\frac{1}{(\alpha+2m)}
(\hbar v_{F}q)^{2m}\nonumber \\
&& \times\Biggl[\frac{1}{E_{q}-\hbar\omega)^{2m+1}}
+\frac{1}{(E_{q}+\hbar\omega)^{2m+1}}\Biggr]
\sum^{m}_{l=0}(-1)^{l}{m\choose l}
\frac{\Gamma\biggl(\frac{\alpha+2l-2}{2}\biggr)}
{\Gamma\biggl(\frac{\alpha+2l}{2}\biggr)}\label{eq:ch}
\end{eqnarray}

\subsection{Imaginary part of the dielectric function} 

The imaginary part of the irreducible polarizability function 
can be derived from Eq.(\ref{eq:pol2}) by using 
Eq.(\ref{eq:identity}) as 
\begin{eqnarray}
Im[\chi^{0}_{\alpha D}(q,\omega)]&=&
-\frac{2^{2-\alpha}}{\pi^{\frac{\alpha-1}{2}}
\Gamma\biggl(\frac{\alpha-1}{2}\biggr)}
\int^{k_{F}}_{0}k^{\alpha-1}dk
\int^{\pi}_{0}\sin^{\alpha-2}\theta d\theta\nonumber \\
&&\times\biggl[\delta\biggl(E_{q}+
\hbar^2kq\cos\theta/m^{\ast}-\hbar\omega\biggr)-
\delta\biggl(E_{q}-\hbar^2kq\cos\theta/m^{\ast}+
\hbar\omega\biggr)\biggr]\label{Imag2}
\end{eqnarray}
Integrating over $\theta$, we find
\begin{eqnarray}
Im[\chi^{0}_{\alpha D}(q,\omega)]=
-\frac{2^{2-\alpha}m^{\ast}}{\pi^{\frac{\alpha-1}{2}}
\Gamma\biggl(\frac{\alpha-1}{2}\biggr)}
&& \biggl[\Theta(k_{F}-k_{-})\int^{k_{F}}_{k_{-}}k^{\alpha-2}dk
\biggl(1-\frac{k^{2}_{-}}{k^{2}}\biggr)^{\frac{\alpha-3}{2}}\nonumber \\
&&\times -\Theta(k_{F}-k_{+})\int^{k_{F}}_{k_{+}}k^{\alpha-2}dk
\biggl(1-\frac{k^{2}_{+}}{k^{2}}\biggr)^{\frac{\alpha-3}{2}}
\biggr]\label{eq:Imag3}
\end{eqnarray}
where $k_{\pm}=\frac{m^{\ast}}{\hbar^{2}q}|\hbar\omega\pm E_{q}|$.

Performing the integration over $k$-space gives the result
\begin{equation}
Im[\chi^{0}_{\alpha D}(q,\omega)]=
-\frac{2^{2-\alpha}m^{\ast}}{(\alpha-1)\pi^{\frac{\alpha-1}{2}}
\Gamma\biggl(\frac{\alpha-1}{2}\biggr)\hbar^{2}q}
\biggl[\Theta(k_{F}-k_{-})(k^{2}_{F}-k^{2}_{-})^{\frac{\alpha-1}{2}}-
\Theta(k_{F}-k_{+})(k^{2}_{F}-k^{2}_{+})^{\frac{\alpha-1}{2}}\biggr]\label{eq:imga4}
\end{equation}

\subsection{Static response function and Plasma dispersion}

The scaled static response function $F_{\alpha D}(q)=
\chi^{(0)}_{\alpha D}(q,\omega\rightarrow 0)/
\chi^{(0)}_{\alpha D}(0,\omega\rightarrow 0)$ 
is a measure of the number of excited states available 
to the system for vanishing excitation energy. Therefore 
$F_{\alpha D}(q,0)$ vanishes in the systems where there 
is a gap in the excitation spectrum. The calculated 
$F_{\alpha D}(q,0)$ for $\alpha=$1, 1.5, 2, 2.5 and 3
at $k_{F}=$0.5 a.u. and $m^{\ast}=m_{0}$ 
are shown in Fig.1.  
According to Eq.(\ref{eq:rs}), the 
$r_{s}$ values corresponding to $k_{F}=$0.5 a.u. 
are $r_{s}$=1.57, 2.25, 2.8,  
3.34 and 3.83 for 
$\alpha$=1, 1.5, 2, 2.5 and 3,
respectively. The static response 
functions in integer dimensions 1D, 2D and 3D  
have been previously reported\cite{Vignale}.   
There are singularities at $q=2k_{F}$ in all dimensions. In 1D there is a 
logarithm divergence. This singular behavior is responsible 
for Peirls instability which is the spontaneous formation of 
density wave at $q=2k_{F}$. In 1.5D the response function is weaker 
and there is a weak kink at $q=2k_{F}$. In 2D the kink at $q=2k_{F}$ 
is quite significant.   
As the dimension is increased, the kink decreases and the derivatives of the 
response functions for $2\le\alpha\le 3$ diverge at $q=2k_{F}$.  
The divergence in response function at $q=2k_{F}$ results in oscillations
with periodicity 2$k_{F}$ in the Fourier transformation of 
$F_{\alpha D}(q,0)$. These are Friedel oscillations which are direct 
consequence of the existence of Fermi surface.  

The plasmon frequency $\omega_{p}(q)$ can be obtained from the roots of 
$\epsilon_{\alpha D}(q,\omega)=0$ which gives the condition, 
\begin{equation}
1-v_{\alpha D}(q)Re[\chi^{0}_{\alpha D}(q,\Omega_{\alpha D}(q))]=
0\label{eq:root}
\end{equation}
Since the analytic solution of this equation does not exist, we find 
$\Omega_{\alpha D}(q)$ by numerical method.  
The plasmon frequencies at $k_{F}$=0.5 a.u. 
for different $\alpha$ values and are shown in Fig.2 . 
The plasma frequency in 3D\cite{Lindhard} has got a finite value at $q=0$.
The plasma frequencies for 2D agree with those of Stern\cite{Stern}. 
The plasma frequency in 1D has been calculated following Das Sarma 
and Hwang\cite{DasSarma}. 
For other dimensions the plasma frequency 
plasma frequency vanishes at $q=0$. 
The condition for the existence of undamped plasma oscillations 
is the the plasma frequency $\Omega_{\alpha D}(q)$ needs to be higher 
than $\omega_{+}(q)=\hbar(2k_{F}q+q^2)/2m^{\ast}$ which 
is the boundary frequency of the  single particle regime. 
The plasma line and e-h line never intersect, but are 
tangential at $q_{c}$. For $q>q_{c}$, the dielectric function has no root
and it is called Landau damping. 
The plasma frequency touches the boundary frequency 
of the single particle regime at $\omega_{+}=\hbar(2k_{F}q+q^2)/2m^{\ast}$

In order to understand this effect,
we evaluate the plasma
frequency in the long-wavelength limit. 
In the long-wavelength limit $q\rightarrow 0$, 
$Im[\chi^{0}_{\alpha D}(q,\omega)]=0$. 
Taking $m=0$ and 1 in the $m$ summation in 
Eq.(\ref{eq:ch}), we find
\begin{equation}
Re[\chi^{0}_{\alpha D}(q,\omega)]=-
\frac{2^{1-\alpha}k^{\alpha}_{F}}{\pi^{\frac{\alpha}{2}}
\Gamma\biggl(1+\frac{\alpha}{2}\biggr)}
\Biggl[\frac{2E_{q}}{E^{2}_{q}-\hbar^{2}\omega^{2}}
+\frac{2\hbar^{2}v^{2}_{F}q^{2}}{\alpha+2}
\biggl(\frac{E^{3}_{q}+3E_{q}\hbar^{2}\omega^{2}}
{(E^{2}_{q}-\hbar^{2}\omega^{2})^{3}}
\biggr)\Biggr]\label{eq:lwl1}
\end{equation}
For $E_{q}<<\hbar\omega$, $(E^{2}_{q}-\hbar^{2}\omega^{2})^{-1}=
-(1+E^{2}_{q}/\hbar^2\omega^2)/\hbar^2\omega^{2}$ and 
$(E^{3}_{q}+3E_{q}\hbar^{2}\omega^{2})/(E^2_{q}-\hbar^{2}\omega^2)^3=
-3E_{q}/\hbar^{4}\omega^{4}$. Substituting these values 
in Eq.(\ref{eq:lwl1}), we find
\begin{equation}
Re[\chi^{0}_{\alpha D}(q,\omega)]=
\frac{2^{1-\alpha}q^{2}k^{\alpha}_{F}}{\pi^{\frac{\alpha}{2}}m^{\ast}
\Gamma\biggl(1+\frac{\alpha}{2}\biggr)\omega^2}
\left[1+\frac{E^{2}_{q}+
3\hbar^2v^{2}_{F}q^2/(\alpha+2)}{\hbar^2\omega^2}\right]\label{eq:lwl2}
\end{equation}
Substituting Eq.(\ref{eq:lwl2}) in Eq.(\ref{eq:root}),
the long-wavelength plasma frequency $\Omega^{lw}_{\alpha D}(q)$ is obtained as 
\begin{equation}
\Omega^{lw}_{\alpha D}(q)=\omega_{\alpha D}(q)
\left[1+\frac{3v^{2}_{F}q^2}
{2(\alpha+2)\omega^{2}_{\alpha D}}+
\frac{\hbar^2q^4}{8m^{\ast}\omega^{2}_{\alpha D}}\right],
\end{equation}
where the classical plasma frequency $\omega_{\alpha D}(q)$ 
is given by\cite{eqn}
\begin{equation}
\omega_{\alpha D}=
\sqrt{\frac{2^{2\alpha-2}\Gamma(1+\frac{\alpha}{2})
\Gamma(\frac{\alpha-1}{2})e^{2}q^{3-\alpha}}
{\sqrt{\pi}\epsilon_{\infty}m^{\ast}r^{\alpha}_{s}}}.\label{eq:clas}
\end{equation}
The dimensionless density parameter is related to the electron density 
$n_{\alpha D}$ as\cite{eqn}
\begin{equation}
r_{s}=\Biggl[\frac{\Gamma(1+\frac{\alpha}{2})}
{\pi^{\frac{\alpha}{2}}n_{\alpha D}}\Biggr]^{\frac{1}{\alpha}}.
\end{equation}

We can easily understand the long-wavelength  $q$-dependence 
in plasma frequency in Fig.1 
by inspecting Eq.(\ref{eq:clas}). The plasma frequency in 3D system is nonzero
at $q=0$ as it independent of $q$ vector. For $\alpha<3$, the plasma frequency 
vanishes at $q=0$. 

\section{Conclusion}

The dielectric function for electron gas in the fractional dimensional 
space has been derived in the RPA. Using the irreducible susceptibilities 
the static response functions for different $\alpha$ values 
have been calculated. The static response functions show derivative 
divergence for all dimensions except for 1D system where 
there is a logarithm singularity. However, the response function 
for 1.5D electron gas is weak.
The plasma dispersion has been found 
from the root of the dielectric function.  
The plasma frequency for low dimensional systems  
vanishes at $q=0$.  It gradually approaches towards bulk value 
when $\alpha$ increases. Ericsson\cite{Plasma2D} experimentally verified the 
plasma frequencies in a wide QW. Similarly the present results 
require experimental verification in a suitable well which shows fractional 
dimensional behavior. In future we are working on local field correction on the dielectric function and plasma frequency using the STLS method.

\begin{figure}[htb]
\includegraphics[width=1\textwidth]{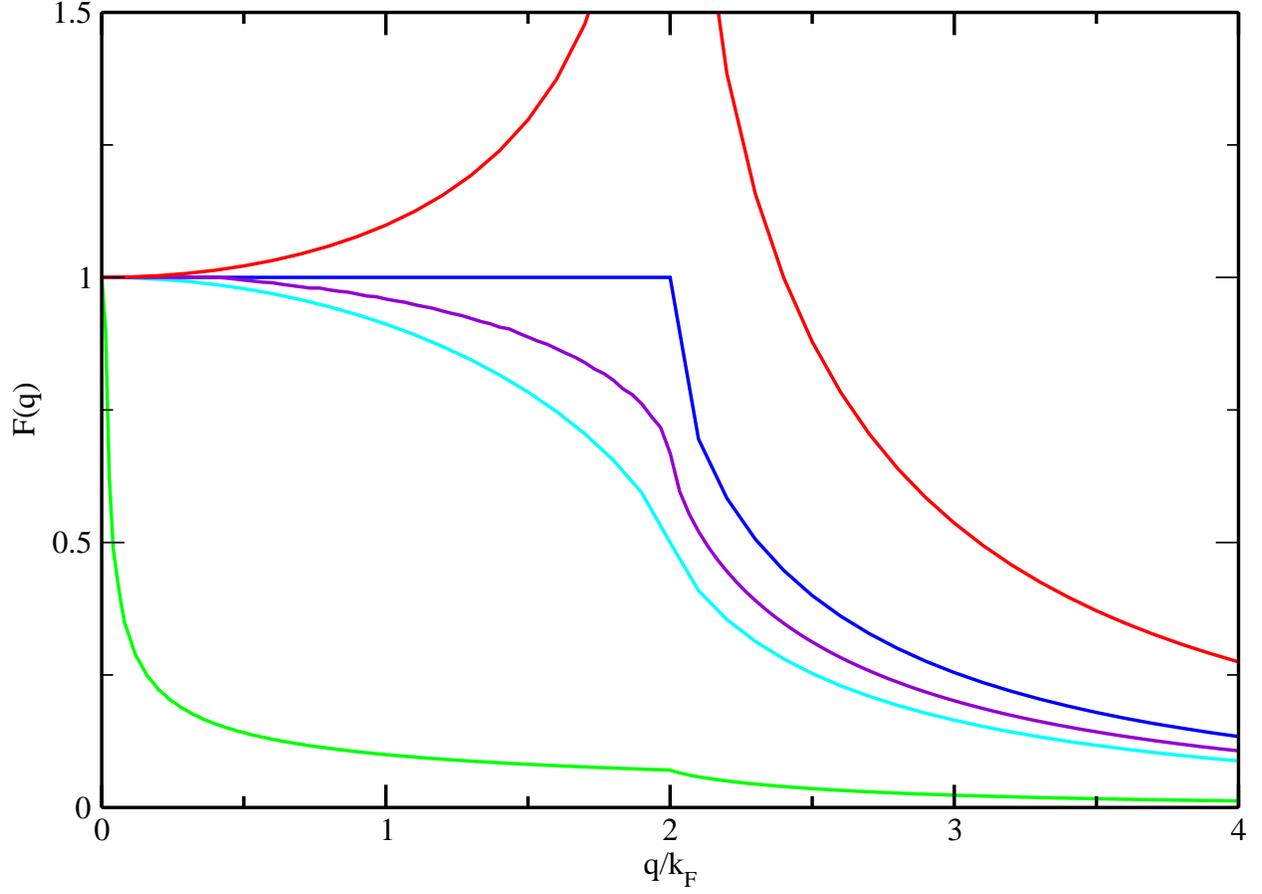}
\caption{Static response functions in fractional dimensions as a function
of wave vector. Here $k_{F}$=0.5 a.u. and $m^{\ast}=m_{0}$. 
Different color lines are denoted as 
$\alpha=$ 1 (red line), 1.5 (green line),  2 (blue line), 2.5 (violet line) and
3 (cyan line).} 
\label{fig:spin1}
\end{figure}
\begin{figure}[htb]
\includegraphics[width=1\textwidth]{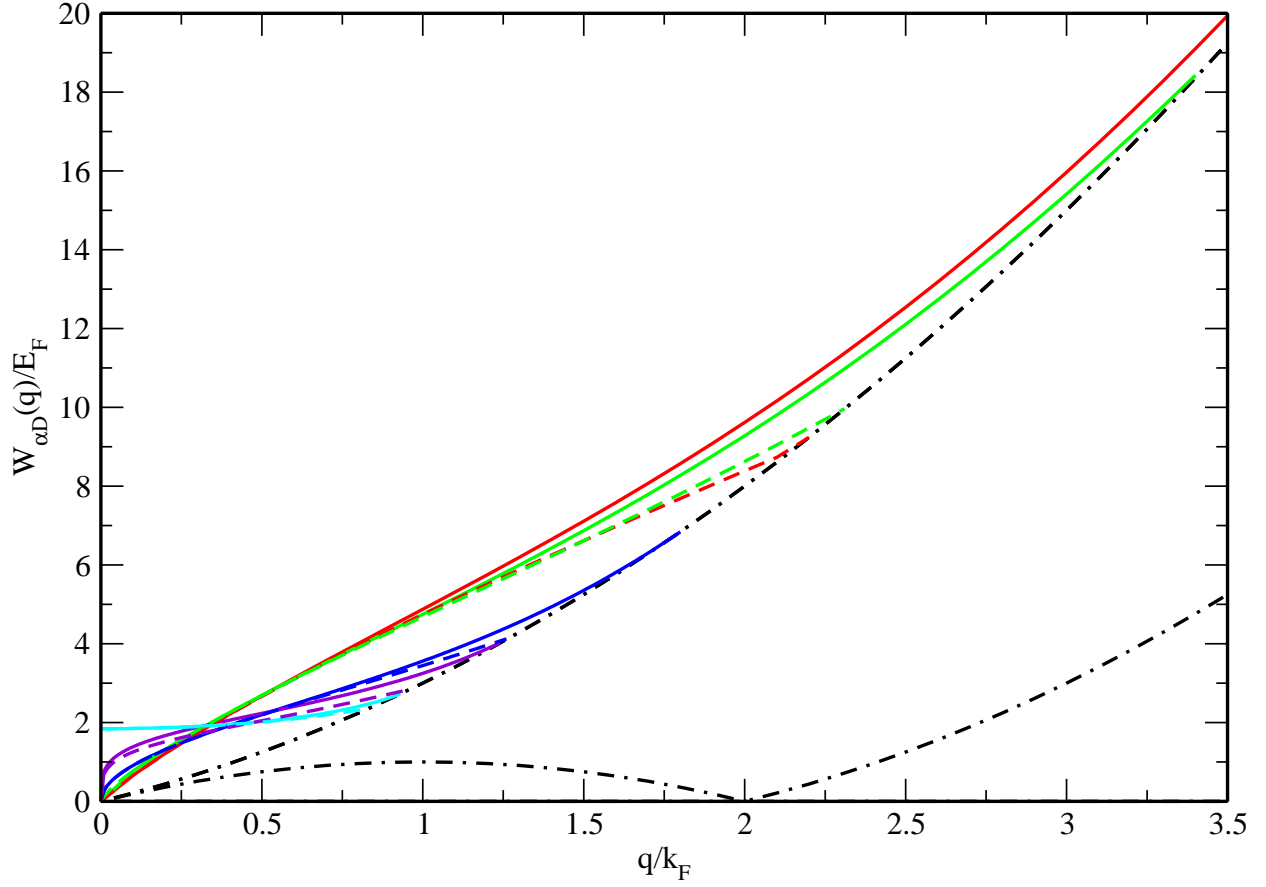}
\caption{Plasma dispersion in fractional dimensions  in several 
dimensions. Here $k_{F}$=0.5 a.u. and $m^{\ast}=m_{0}$. 
The black dot-dashed line represents the boundary of the 
single particle regime. Different color lines correspond to different 
dimensions as in Fig.1. Plasma frequencies are denoted by solid lines 
while plasma frequencies at long-wavelength limit are denoted by 
dashed lines.} 
\label{fig:spin2}
\end{figure}

\begin{references}
\bibitem{eqn}Standard expression in literature is $\omega_{\alpha D}=
\sqrt{(4\pi)^{{{\alpha-1}\over 2}}\Gamma({{\alpha-1}\over 2})e^2n_{\alpha D}
q^{3-\alpha}/\epsilon_{\infty}m^{\ast}}$, where high-frequency 
dielectric constant $\epsilon_{\infty}$ is taken unity.
\end{references}
\end{document}